\documentclass[11pt,nofootinbib,aps,preprint]{revtex4-1}

\usepackage{amsmath,mathrsfs}
\usepackage{graphicx,epsfig}
\usepackage{latexsym,amssymb}
\usepackage{epsf}
\usepackage{ifpdf,natbib,lineno}

\usepackage{xcolor}

\begin{document}

\title{{\bf  Zero Rest Mass Soliton Solutions}}

\author{ M. Mohammadi$^{1}$ } \email{physmohammadi@pgu.ac.ir}   \author{R. Gheisari$^{1,2}$} \email{gheisari@pgu.ac.ir} \address{$^1$Physics Department, Persian Gulf University, Bushehr, 75169, Iran\\ $^2$ Nuclear Energy Research Center, Persian Gulf University, Bushehr, 75169, Iran}


\begin{abstract}

In this paper, extended Klein-Gordon field systems will be introduced. Theoretically, it will be shown that for a special example of these systems, it is possible to have a single zero rest mass soliton solution, which is forced to move at the speed of light provided it is considered a non-deformed rigid object. This special soliton solution has the minimum energy among the other solutions, i.e. any arbitrary deformation in its internal structure leads to an increase in the total energy.


\end{abstract}

\maketitle

 \textbf{Keywords} : {zero rest mass; soliton; stability; extended Klein-Gordon; solitary wave}

\section{Introduction}

The classical relativistic field theory with soliton solutions is an attempt to model particles as some stable localized non-singular objects. Solitons are the stable solitary wave solutions of the nonlinear field systems  whose energy density functions are localized and satisfies the standard relativistic energy-rest mass-momentum relations properly \cite{rajarama,Das,lamb,Drazin}.  A well-known example for solitons in $1 + 1$ dimensions are kink and anti-kink solutions of the real nonlinear Klein-Gordon (KG) systems \cite{phi41,phi42,phi43,phi44,phi45,OV,GH,MM1,MR,phi62,DSG1,DSG2,DSG3,MM2,waz,Kink1,Kink2,Kink3,Kink4,Kink5,Kink6,Kink7}. For the complex nonlinear KG systems, it was shown that there are some localized soliton (-like) solutions that are called Q-balls \cite{Vak3,Vak4,Vak5,Vak6,Lee3,Scoleman,R1,R2,R3,R4,R5,R6,R7,R8,Riazi2,MM3}. Morover, the Skyrme model  \cite{SKrme,SKrme2} and 't Hooft Polyakov model  \cite{toft,pol} are also two well-known relativistic field models, which yield soliton solutions in $3+1$ dimensions.

All well-known relativistic linear and  nonlinear scalar field  systems,  which have been used  in the classical or quantum field   theory,  are in the standard mathematical formats  that  can be called  (nonlinear) KG (-like) systems. In this work,  the extended  KG systems   for scalar fields are introduced as well. Theoretically, we show that for a special  extended  KG system in $1+1$ dimensions,  as a simple  example, it is possible to have a  soliton solution with zero rest mass (or zero energy). The other non-trivial solutions of this special system  all have  non-zero  positive rest energies. In fact,  the   massless soliton solution is a  special single solution among the others with the least amount of energy.  This massless solution, if it is considered  a non-deformed rigid object,  responds to any amount
of force, no matter how tiny, by immediately accelerating to the speed of light. This  model can be simply extended to $3+1$ dimensions to have a particle-like solution with zero rest mass. In general, if one considers such a massless particle in the space, since there is no area in the space with absolute zero interaction, it has to move at the speed of light, exactly according to the special relativity.

The organization of this paper is as follows. In section 2, the (nonlinear) KG systems and the extended KG systems are introduced. In section 3, a special extended KG system is introduced, which leads to a stable massless soliton solution. The last section is devoted to the summary and conclusions.

\section{The (nonlinear) KG systems and the extended KG  systems}\label{sec2}

In the standard relativistic   field theory, different systems can be identified with different Lagrangian densities. The standard  Lagrangian densities  are considered to be functions of several  fields $\phi_{r}$ ($r=1,\ldots,N$) and their first  derivatives ($\phi_{r,\alpha}=\frac{\partial\phi_{r}}{\partial x^{\alpha}}$):
 \begin{equation} \label{lag}
{\cal L}= {\cal L}(\phi_{r},\phi_{r,\alpha}).
 \end{equation}
According to the principle of least action, the related equations of motion would be,
 \begin{equation} \label{eqomo}
\frac{\partial{\cal L}}{\partial\phi_{r}}-\frac{\partial}{\partial x^{\mu}}\left(\frac{\partial {\cal L}}{\partial \phi_{r,\mu}}\right)=0.
 \end{equation}
In general, since the Lagrangian density (\ref{lag}) is invariance  under space-time translations $x^{\mu}\rightarrow x^{\mu}+a^{\mu}$, four continuity equations $\partial_{\mu}T^{\mu\nu}=0$ and hence four conserved quantities $P^{\mu}=\int T^{o\mu}d^{3}\textbf{x}$ are obtained where,
 \begin{equation} \label{EMT}
T^{\mu\nu}=\frac{\partial{\cal L}}{\partial\phi_{r,\mu}}\frac{\partial\phi_{r}}{\partial x_{\nu}}-{\cal L}g^{\mu\nu}
\end{equation}
is called the energy-momentum tensor and $g^{\mu\nu}$     the $3 + 1$ dimensional Minkowski metric. Note that, in this paper, for simplicity, we take the speed of light equal to one ($c=1$). For any  arbitrary solution  for which the   components  of the  energy-momentum tensor $T^{\mu\nu}$ asymptomatically approach  zero at infinity, the four conserved quantities $P^{\mu}=\int T^{o\mu}d^{3}\textbf{x}$ form  a four vector, which is called the energy-momentum four vector. So, for all localized solutions of equation (\ref{eqomo}), the standard relativistic  energy-rest mass-momentum  relations  satisfied generally:
\begin{eqnarray} \label{d}
&&E=m =P^{0} \equiv \int T^{00} d^{3}\textbf{x}=\gamma E_{o}=\gamma m_{o}, \\&&
\textbf{P}=(P^{1},P^{2},P^{3})=\int \left(T^{01},T^{02},T^{03} \right)d^{3}\textbf{x}=\gamma m_{o} \textbf{v}.
\end{eqnarray}
Note that, $E_{o}$ ($m_{o}$) is the rest energy (mass) for a non-moving localized solution and $E$ ($m$) is  the total energy (mass) for its moving version.

Many of the known standard Lagrangian densities, which are used in quantum   or classical relativistic field  theory,  are in the same formats that  can be called (nonlinear) KG (-like)  Lagrangian densities. The (nonlinear) KG Lagrangian densities can be defined as the linear functions of the kinetic  scalar terms ${\cal S}_{ij}={\cal S}_{ji}=\partial_{\mu}\phi_{i}\partial^{\mu}\phi_{j}$, i.e.
 \begin{equation} \label{asd}
{\cal L}= \sum_{j\geqslant i}^{N}\sum_{i}^{N} \alpha_{ij}(\phi_{1},\cdots,\phi_{N}){\cal S}_{ij}-V(\phi_{1},\cdots,\phi_{N}),
 \end{equation}
where $\phi_{i}$ must be some scalar fields (not a vector field $\phi_{i\mu}$, like electromagnetic vector field $A^{\mu}$),  $V$ (field potential) and coefficients $\alpha_{ij}$'s all are functions of the scalar fields.     For example, if one considers a  complex scalar field $\phi_{1}=\phi$ ($\phi_{2}=\phi^{*}$ ), the allowed kinetic scalar terms are ${\cal S}_{11}=\partial_\mu \phi\partial^\mu \phi$, ${\cal S}_{22}=\partial_\mu \phi^*\partial^\mu \phi^*$ and ${\cal S}_{12}=\partial_\mu \phi\partial^\mu \phi^*$. Note that the  permissible  combinations of  the kinetic  scalars must be ones for which the constructed     Lagrangian density (\ref{asd}) is a   real functional. For example, the  known complex nonlinear KG  Lagrangian densities,  which yield non-topological particle-like solutions (Q-balls) \cite{Vak3,Vak4,Vak5,Vak6,Lee3,Scoleman,R1,R2,R3,R4,R5,R6,R7,R8,Riazi2,MM3},  are introduced as  follows:
 \begin{equation} \label{exam}
{\cal L}= \partial_\mu \phi^*
\partial^\mu \phi -V(|\phi |)={\cal S}_{11}-V(|\phi |),
 \end{equation}
for which $\alpha_{11}=\alpha_{22}=0$ and $\alpha_{12}=1$. Equivalently, one can decompose  complex scalar field $\phi$ into two distinct real and imaginary parts $\phi_{1}=\phi_{r}$  and $\phi_{2}=\phi_{i}$ (i.e. $\phi=\phi_{r}+i\phi_{i}$).  In this representation, the allowed kinetic scalar terms are  ${\cal S}_{11}=\partial_{\mu}\phi_{r}\partial^{\mu}\phi_{r}$, ${\cal S}_{22}=\partial_{\mu}\phi_{i}\partial^{\mu}\phi_{i}$ and ${\cal S}_{12}=\partial_{\mu}\phi_{r}\partial^{\mu}\phi_{i}$, respectively. Therefore, the Lagrangian density (\ref{exam}) becomes,
\begin{equation} \label{cnk}
{\cal L}=\partial_\mu \phi_{r}
\partial^\mu \phi_{r} +\partial_\mu \phi_{i}
\partial^\mu \phi_{i}-V(\sqrt{\phi_{r}^2+\phi_{i}^2}),
\end{equation}
for which $\alpha_{11}=\alpha_{22}=1$ and $\alpha_{12}=0$. Moreover, instead of scalars fields  $\phi$ and $\phi^{*}$, one can change the variables to the polar fields $R(x^{\mu})$ and $\theta(x^{\mu})$ as defined by,
\begin{equation} \label{polar}
 \phi(x,y,z,t)= R(x,y,z,t)\exp[i\theta(x,y,z,t)].
\end{equation}
In the representation of polar fields, $\phi_{1}=R=|\phi |=\sqrt{\phi\phi^{*}}$ is the module scalar field   and $\phi_{2}=\theta$ is the phase scalar field, while  the allowed kinetic  scalar terms    are  ${\cal S}_{11}=\partial_{\mu}R\partial^{\mu}R$, ${\cal S}_{22}=\partial_{\mu}\theta\partial^{\mu}\theta$ and ${\cal S}_{12}=\partial_{\mu}R\partial^{\mu}\theta$. In terms of  polar fields, the Lagrangian-density (\ref{exam}) becomes,
\begin{equation} \label{cnk}
{\cal L}=(\partial^\mu R\partial_\mu R) +R^{2}(\partial^\mu\theta\partial_\mu\theta)-V(R)=({\cal S}_{11}) +R^{2}({\cal S}_{22})-V(R),
\end{equation}
for which $\alpha_{11}=1$, $\alpha_{22}=R^2$  and $\alpha_{12}=0$.  Therefore, due to different equivalent fields (representations), there are different kinetic scalars and one can use them to introduce any arbitrary  new (nonlinear) KG  system.

As previously mentioned, the (nonlinear) KG systems are linear functions of the kinetic scalar terms ${\cal S}_{ij}$ (\ref{asd}). It is possible to  consider  systems, which are not linear in terms of the kinetic scalars  ${\cal S}_{ij}$. We can call them  "\emph{extended  KG systems}", which are essentially nonlinear.
   Namely, for a complex field in the polar representation, i.e.  $\phi_{1}=R$ and $\phi_{2}=\theta$, the following Lagrangian densities  are  some examples of the extended  KG systems:
\begin{eqnarray} \label{sdc}
&&{\cal L}=(\partial^\mu R\partial_\mu R)(\partial^\mu\theta\partial_\mu\theta)-V(R)=({\cal S}_{11})({\cal S}_{22})-V(R),\\&&
{\cal L}=(\partial^\mu R\partial_\mu R)+R^2(\partial^\mu\theta\partial_\mu\theta)^2-V(R)=({\cal S}_{11})+R^2({\cal S}_{22})^2-V(R),\\&&
{\cal L}=\left[(\partial^\mu R\partial_\mu R)+R^2(\partial^\mu\theta\partial_\mu\theta)-V(R)\right]^2=\left[({\cal S}_{11})+R^2({\cal S}_{22})-V(R)\right]^2\\&&
{\cal L}=\left[(\partial^\mu R\partial_\mu R)+R^2(\partial^\mu\theta\partial_\mu\theta)\right]^2-V(R)=\left[({\cal S}_{11})+R^2({\cal S}_{22})\right]^2-V(R)
\end{eqnarray}
For an extended one,  it is expected to encounter  very complicated equations and other  properties.  Although it is not usual to use  extended  KG systems   in  standard (quantum or classical) field theory,  in this paper it will be shown that using some special extended  KG Lagrangian densities   can  lead to some interesting particle-like solutions with zero rest masses.


\section{An extended  KG system with a single massless  soliton solution in $1+1$ dimensions}\label{sec4}

In this section, we are going to show that it is  possible to have an extended  KG system in $1+1$ dimensions  with a single non-trivial massless  stable particle-like solution. A similar  model can be introduced in $3+1$ dimensions. However for simplicity,  we restrict ourself in $1+1$ dimensions.  To achieve this goal,  we can consider  an extended  KG system for a single complex scalar field $\phi~(=R e^{i\theta})$, or equivalently for two independent  scalar fields  $\phi_{1}=R$ and  $\phi_{2}=\theta$, in the following form:
 \begin{equation} \label{kexn}
{\cal L}=\sum_{i=1}^3{\cal K}_{i}^3,
 \end{equation}
where
 \begin{eqnarray} \label{e5}
&&{\cal K}_{1}=R^2[{\cal S}_{22}-2],\\&&\label{e52}
{\cal K}_{2}=R^2[{\cal S}_{22}-2]+[{\cal S}_{11}-4R^4+4R^3], \\&&\label{e53}
{\cal K}_{3}=R^2[{\cal S}_{22}-2]+[{\cal S}_{11}-4R^4+4R^3]+2 R[{\cal S}_{12}],
\end{eqnarray}
in which,  ${\cal S}_{11}=\partial_{\mu}R\partial^{\mu}R$, ${\cal S}_{22}=\partial_{\mu}\theta\partial^{\mu}\theta$, ${\cal S}_{12}=\partial_{\mu}R\partial^{\mu}\theta$. The related equations of motion can be obtained easily:
\begin{eqnarray} \label{jkt}
&&\sum_{i=1}^3 {\cal K}_{i}\left[2(\partial_{\mu}{\cal K}_{i})   \frac{\partial{\cal K}_{i}}{\partial(\partial_{\mu}R)}    +   {\cal K}_{i}\partial_{\mu}\left(\frac{\partial{\cal K}_{i}}{\partial(\partial_{\mu}R)}\right)    -    {\cal K}_{i}\frac{\partial{\cal K}_{i}}{\partial R}   \right]=0,\\&&\label{jkt2}
\sum_{i=1}^3 {\cal K}_{i}\left[2(\partial_{\mu}{\cal K}_{i})   \frac{\partial{\cal K}_{i}}{\partial(\partial_{\mu}\theta)}    +   {\cal K}_{i}\partial_{\mu}\left(\frac{\partial{\cal K}_{i}}{\partial(\partial_{\mu}\theta)}\right)       \right]=0.
\end{eqnarray}
The energy density that belongs to the new extended Lagrangian density (\ref{kexn}),   would be,
\begin{eqnarray} \label{MTEr}
&&\varepsilon(x,t)=T^{00}=\sum_{i=1}^{3}{\cal K}_{i}^{2}\left[3C_{i}
-{\cal K}_{i}\right]=\varepsilon_{1}+\varepsilon_{2}+\varepsilon_{3},
\end{eqnarray}
which is divided  into three  distinct  parts, in which,
\begin{equation}\label{cof}
C_{i}=\dfrac{\partial{\cal K}_{i}}{\partial \dot{\theta}}\dot{\theta}+\dfrac{\partial{\cal K}_{i}}{\partial \dot{R}}\dot{R}=
\begin{cases}
\quad\quad 2R^2\dot{\theta}^{2} & \text{i=1}
\\
2(\dot{R}^{2}+R^2\dot{\theta}^2) & \text{i=2}
\\2(\dot{R}+R\dot{\theta})^2
 & \text{i=3}.
\end{cases}
\end{equation}
After a straightforward calculation, we obtain:
 \begin{eqnarray} \label{eis}
&&\varepsilon_{1}={\cal K}_{1}^2[5R^2\dot{\theta}^2+R^2\theta'^2+2R^2],\\ \label{eis1}&&
\varepsilon_{2}={\cal K}_{2}^2[5R^2\dot{\theta}^2+5\dot{R}^2+R^2\theta'^2+R'^2+U(R)], \\ \label{eis2}&&
\varepsilon_{3}={\cal K}_{3}^2[5(R\dot{\theta}+\dot{R})^2+(R\theta'+R')^2+U(R)],
\end{eqnarray}
where the dot and prime denote differentiation with respect to $t$ and $x$ respectively, and
\begin{eqnarray} \label{UR}
U(R)=4R^4-4R^3+2R^2,
\end{eqnarray}
is an  ascending function, which is bounded from below by zero. Therefore,  all terms in equations (\ref{eis}), (\ref{eis1}) and (\ref{eis2}) are positive definite and then the  energy density function (\ref{MTEr}) is bounded from below by zero too. Hence, at any arbitrary space-time point, the possible  minimum value of the energy density function (\ref{MTEr}) is zero. For example, for the trivial  solution $R=0$, i.e. the vacuum state, the energy density function would be zero everywhere. Moreover, there is a single  non-trivial localized solution for which the energy density function (\ref{MTEr}) is zero everywhere. In fact, it is a non-trivial localized solitary wave solution with zero energy (rest mass) as well.

In general, according to  equations (\ref{jkt}), (\ref{jkt2}), (\ref{eis}), (\ref{eis1}) and (\ref{eis2}), it is obvious  that the special solutions with zero (rest) energies   are ones   for which ${\cal K}_{i}$ ($i=1,2,3$) all are zero simultaneously. Therefore, to obtain such special solutions, in general,  three  conditions ${\cal K}_{i}=0$ ($i=1,2,3$) must be satisfied simultaneously, which  lead to  three independent nonlinear  partial differential equations (PDEs) for two independent scalar  fields $R$ and $\theta$ as follows:
 \begin{eqnarray} \label{dfb}
&&\partial_{\mu}\theta\partial^{\mu}\theta=2,\\&& \label{dfb2}
\partial_{\mu}R\partial^{\mu}R=4R^4-4R^3, \\&& \label{dfb3}
\partial_{\mu}R\partial^{\mu}\theta=0,
\end{eqnarray}
which do not have any common  solution except  a trivial solution $R=0$  and a single non-trivial solitary wave  solution   in the following form:
\begin{equation} \label{So}
R_{s}(x)=\dfrac{1}{1+x^2},\quad  \quad\theta_{s}(t)=\omega_{s}t,
\end{equation}
which is at rest and $\omega_{s}=\pm\sqrt{2}$ can be called  "\emph{the rest frequency}". Using the  Lorentz transformations, the moving version of this special solution (\ref{So}) can be obtained easily as follows:
\begin{equation} \label{So2}
R_{v}(x,t)=\dfrac{1}{1+\gamma^2(x-vt)^2},\quad  \quad\theta_{v}(x,t)=k_{\mu}x^{\mu},
\end{equation}
in which $v$ is the velocity,  $\gamma=1/\sqrt{1-v^2}$ and  $k^{\mu}\equiv(\omega,k)$ is a $1+1$ vector, provided
\begin{equation} \label{pro}
k= {\omega}v,
\end{equation}
and
\begin{equation} \label{pro2}
\omega=\gamma\omega_{s}.
\end{equation}

\begin{figure}[ht!]
   \centering
   \includegraphics[width=160mm]{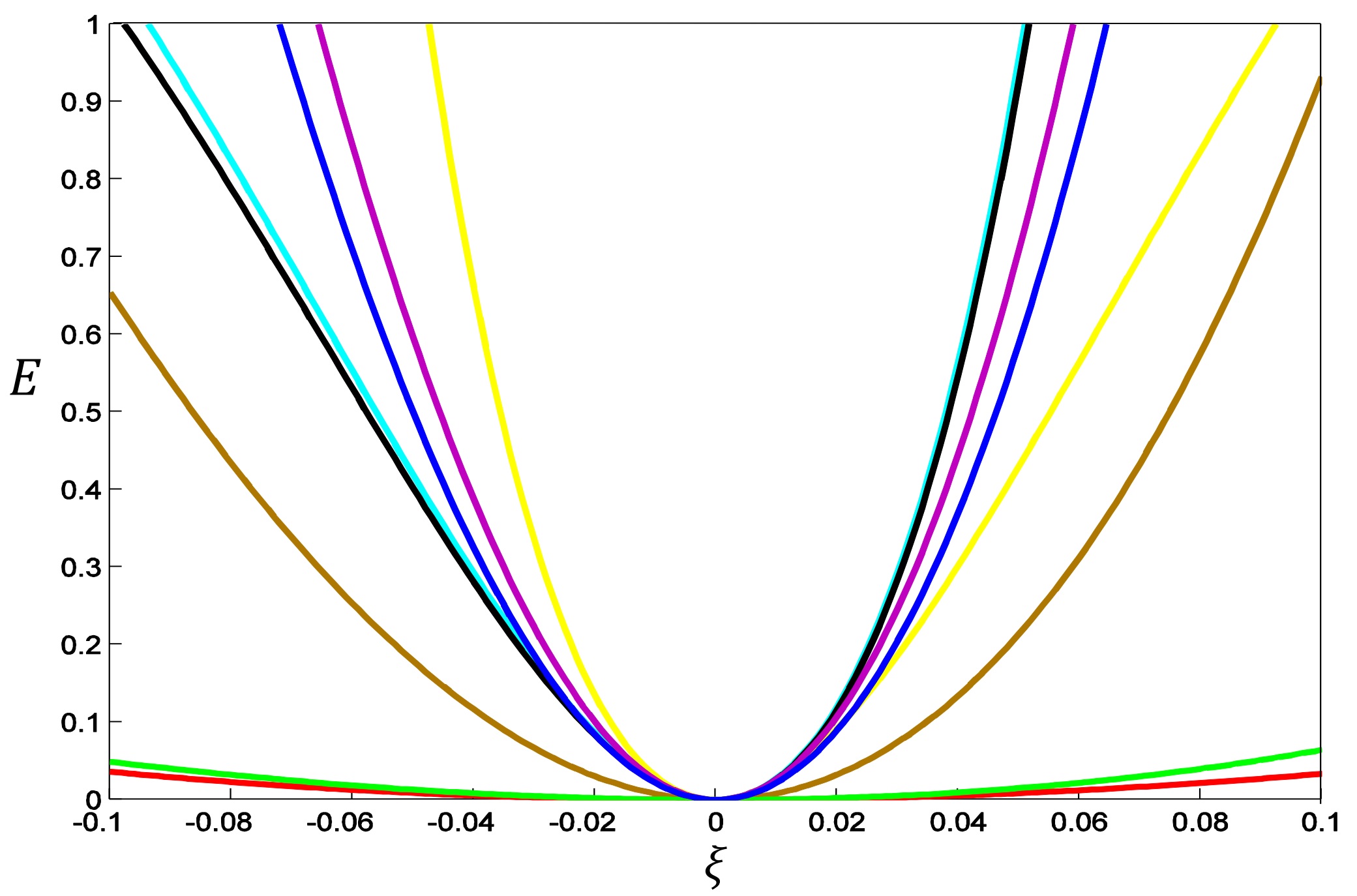}
   \caption{Variations  of  the total  energy $E$ versus small $\xi$  for  different deformations (\ref{var1})-(\ref{var8}) at $t=0$.  Various color curves of cyan, black, red, yellow, brown, purple, green and blue are related to various deformations (\ref{var1})-(\ref{var8}), respectively.} \label{1}
\end{figure}

Since the energy density function  (\ref{MTEr})  is    positive definite, therefore the  special  solution (\ref{So}), for which ${\cal K}_{i}=0$ ($i=1,2,3$),  definitely has  the minimum rest energy ($E_{o}=0$) among the other solutions, i.e. it is a soliton solution.
More precisely,  for any non-trivial solution of the  equations (\ref{jkt}) and (\ref{jkt2}), except the special solution (\ref{So}), three independent conditions (\ref{dfb}), (\ref{dfb2}) and (\ref{dfb3}),  as three independent PDEs, are not possible to satisfied simulatively. Therefore,  for other solutions of the dynamical equations (\ref{jkt}) and (\ref{jkt2}),      at least one of the independent  scalars ${\cal K}_{i}$ takes non-zero value, which leads to a non-zero positive energy density  function (see equations (\ref{eis}), (\ref{eis1}) and (\ref{eis2})), i.e. the rest energy is zero just for the single solitary wave solution (\ref{So}), and for other non-trivial unknown solutions lead to non-zero positive values. In other words,   for any arbitrary  deformation above the background of the  special solution (\ref{So}) (i.e. $R=R_{s}+\delta R$ and $\theta=\theta_{s}+\delta\theta$), the total energy always increases. For example, for eight different arbitrary small  deformations which  are introduced as follows:
\begin{equation} \label{var1}
R=\dfrac{1}{1+x^2}+\xi e^{-x^2},\quad  \quad\theta=\omega_{s}t,
\end{equation}
\begin{equation} \label{var2}
R=\dfrac{(1+\xi)}{1+x^2},\quad  \quad\theta=\omega_{s}t,
\end{equation}
\begin{equation} \label{var3}
R=\dfrac{1}{1+(1+\xi)x^2},\quad  \quad\theta=\omega_{s}t,
\end{equation}
\begin{equation} \label{var4}
R=\dfrac{1}{(1+\xi)+x^2},\quad  \quad\theta=\omega_{s}t,
\end{equation}
\begin{equation} \label{var5}
R=\dfrac{1}{1+x^2}+\xi t e^{-x^2},\quad  \quad\theta=\omega_{s}t,
\end{equation}
\begin{equation} \label{var6}
R=\dfrac{1}{1+x^2},\quad  \quad\theta=\omega_{s}t+\xi t,
\end{equation}
\begin{equation} \label{var7}
R=\dfrac{1}{1+x^2},\quad  \quad\theta=\omega_{s}t+\xi e^{-x^2},
\end{equation}
\begin{equation} \label{var8}
R=\dfrac{1}{1+x^2},\quad  \quad\theta=\omega_{s}t+\xi t e^{-x^2},
\end{equation}
where it is easy numerically to obtain the curves  of  the total energy $E$ versus $\xi$ (see Fig.~\ref{1}). Here, $\xi$ is a small parameter and can be considered as  an  indication  of  the order  of   deformations. Note that,  the case $\xi=0$  leads to the same special solitary wave solution (\ref{So}). Figure \ref{1}, just as some examples,  properly shows how the special solution (\ref{So}) is stable against any arbitrary deformation.

 Due to relativity, no particle can move faster than the speed
of light, and therefore this massless particle (\ref{So}), if it is considered  a rigid
body,  will respond to any amount of force (interaction), no matter how tiny, by immediately accelerating to the speed of light. But, since the special  solution (\ref{So})  is  a composite   of two deformable fields $R$ and $\theta$ (as its internal structure), and then is not essentially a rigid body, it would   be   deformed ($R=R_{s}+\delta R$ and $\theta=\theta_{s}+\delta\theta$)   in  collisions  or in interactions with other particles. In other words, it is not actually  possible to perceive it as a rigid object with absolute zero rest mass.
In fact, Heisenberg's uncertainty principle essentially does not  let us  to have an  ideal  state for which the energy is absolute zero, and this statement is also valid  for the special particle-like solution (\ref{So}). Hence, physically, the particle-like solution (\ref{So}) is always slightly deformed and then its rest mass  is not absolute zero but it is very small, i.e. it is a stable soliton  solution, which    moves at a speed very close to the speed of light, but not exactly equal to the speed of light.



To  establish  an extended  KG  model  (\ref{kexn}) for which   the stability  of  the single non-trivial solitary wave solution (\ref{So}) is  guaranteed appreciably through a simple and straightforward  conclusion, we select three independent   linear combinations of ${\cal S}_{ij}$'s in relations (\ref{e5}), (\ref{e52}) and (\ref{e53}) for which the functional coefficients $C_{i}$ ($i=1,2,3$) would be definitely positive and finally lead to a  positive definite energy density function (\ref{MTEr}).   In general,  it may be possible to choose other  combinations of ${\cal S}_{ij}$ for this goal. However, we intentionally introduced these  special combinations   as a good example of the extended   KG systems for better and simpler  conclusions.

\section{Summary and conclusion}\label{sec4}


We have introduced, after reviewing some basic relations and equations of the standard relativistic classical field
theory, the (nonlinear) KG and the extended KG systems for the scalar fields $\phi_{i}$.  The Lagrangian density of a (nonlinear) KG system   is linear in  the kinetic scalar terms ${\cal S}_{ij}={\cal S}_{ji}=\partial_{\mu}\phi_{i}\partial^{\mu}\phi_{j}$.  If the Lagrangian density of the scalar fields is not linear in the  kinetic scalar terms, it can be called an extended  KG system.

In this paper, it was shown that for a special example of the extended  KG systems (\ref{kexn}), there is a single  non-trivial localized solitary wave solution with zero rest mass (\ref{So}). The   energy density function of this special extended  KG system is bounded from below by zero. Therefore, the single solitary wave solution (\ref{So}) has the minimum energy among the other solutions, i.e. it is really a massless  stable localized  solution. In other words, it is a zero rest mass soliton solution. The other unknown solutions of this system, undoubtedly, have non-zero positive rest energies.

The single  massless soliton solution (\ref{So}), if  considered  a rigid body, responds to any amount of force,
no matter how tiny, by immediately accelerating to the speed of light. Therefore, since there is no  area  in the space with  zero  interaction, it  has to move at the speed of light exactly according to the special relativity. But, since the stable
zero rest mass solution (\ref{So}) has internal structure and is not essentially a rigid body, it would be
deformed in collisions or in interactions with other particles. Hence, it is not physically possible
to perceive  it as an absolute zero rest energy (mass) object. In fact, the zero
rest mass soliton solution  (\ref{So})  is just an  ideal mathematical solution, which  can be considered in a free space without any other particles
and interactions, but  is not an interesting physical case. Hence, physically,  the deformed soliton solution   (\ref{So}) can move at a speed very close to the speed of light, but  not exactly equal to the speed of light.

%
%
%
%

\section*{ACKNOWLEDGEMENT}

The authors acknowledge the Persian Gulf University Research Council.

\end{document}